\documentclass[onecollarge,natbib]{svjour2}
\bibpunct{[}{]}{,}{n}{}{,} % to get "[numbered]" references from natbib
\smartqed  % flush right qed marks, e.g. at end of proof
\usepackage{graphicx}
\usepackage{epsfig}
\def\be{\begin{equation}} \def\ee{\end{equation}} \def\bea{\begin{eqnarray}}
\def\eea{\end{eqnarray}} \def\nnb{\nonumber}
\newcommand{\no}{\nonumber \\}

\journalname{Few-Body Systems (APFB2011)}
\begin{document}

\title{\boldmath
Spin observables in the two-nucleon capture and dissociation processes
at low energies}

\author{Y.-H. Song\and C. H. Hyun \and S.-I. Ando \and K. Kubodera}


\institute{Y.-H. Song, K. Kubodera \at
Department of Physics and Astronomy,
University of South Carolina, Columbia, SC 29208, USA
\and
C. H. Hyun, S.-I. Ando \at
Department of Physics Education,
Daegu University, Gyeongsan 712-714, Republic of Korea\\
\email{hch@daegu.ac.kr}
}

\date{Received: date / Accepted: date}

\maketitle

\begin{abstract}
Spin observables in radiative neutron capture on a proton
and its inverse process, photodisintegration of the deuteron
are calculated using a pionless effective field theory with di-baryon fields.
Good agreement with the results of existing standard nuclear 
physics approach is obtained at very low energies.
As energy increases, however, the discrepancy between the effective
field theory and the standard nuclear physics approach becomes substantial. 
We discuss the origin of the difference.

\end{abstract}

\section{Introduction}
\label{intro}

The importance of spin observables in nuclear physics is in the fact 
that they can give more detailed information on the dynamics of the system.
Because spin observables are sensitive to the transition amplitudes,
they can be good testing grounds to check the accuracy of a theory.
Traditional standard nuclear physics approach(SNPA) uses phenomenological
potential models and current operators which satisfy the current conservation.
SNPA could explain many nuclear phenomena including total cross sections
of radiative neutron capture on a proton. 
Recently effective field theory(EFT) approach has been widely used. 
EFT approach is model-independent, provides theoretical error estimation,
and its accuracy can be improved in a systematic way.
EFT calculations showed good agreement with 
experiments and SNPA calculations at low energies with relatively
less number of parameters for many observables.
However, most of the calculations are focused on unpolarized observables
which are dominated by a few partial wave amplitudes. 
It would be interesting to check whether the same amplitudes 
can explain spin observables which are sensitive to the interference
between the amplitudes.  

It was observed that there exist discrepancies between SNPA calculation and
experiments in the induced neutron polarization, $P_{y'}$, 
in photo-disintegration of the deuteron \cite{rocco2005}. 
Recent pionless di-baryon effective field theory(dEFT) calculation
up to next-to-leading order showed good agreement with SNPA at low energies, 
but showed conspicuous difference with both theoretical SNPA calculation
and measurements at energies larger than 8 MeV \cite{ashk2011}. 
Though it is not yet clear
whether the differences in experiments and theory are genuine or not,
it is important to check whether the same amplitudes used for $P_{y'}$ 
can explain other spin observables and whether the difference between SNPA
and dEFT at high energies can be explained by increasing the accuracy of dEFT. 

In this work, we calculate the left-right asymmetry in the polarized
neutron capture on a proton, and the linear polarization of the photon
in the disintegration of the deutron at low energies using the dEFT.
We find good agreement with the results obtained from SNPA at low
energies, but the discrepancy becomes significant as the energy increases.
\section{Formalism}
\label{formalism}

We relegate details of the calculation to Ref.~\cite{ashk2011}
and only summarize the results.
Relevant transition amplitude in the c.m. frame for $\gamma d\to n p$ is
\bea
A &=&
\chi_1^\dagger \vec{\sigma}\sigma_2\tau_2\chi_2^{T\dagger}\cdot
\left\{
[\vec{\epsilon}_{(d)}\times(\hat{k}\times\vec{\epsilon}_{(\gamma)})] X_{MS}
+ \vec{\epsilon}_{(d)}\vec{\epsilon}_{(\gamma)}\cdot\hat{p}\, Y_{ES}
\right\}
\nnb \\ &&
+ \chi_1^\dagger \sigma_2\tau_3\tau_2\chi_2^{T\dagger}
i\vec{\epsilon}_{(d)}\cdot (\hat{k}\times\vec{\epsilon}_{(\gamma)})\,
X_{MV}
\nnb \\ &&
+ \chi_1^\dagger \vec{\sigma}\sigma_2\tau_3\tau_2\chi_2^{T\dagger} \cdot
\left\{
\vec{\epsilon}_{(d)}\vec{\epsilon}_{(\gamma)}\cdot\hat{p}\, X_{EV}
+ [\vec{\epsilon}_{(d)}\times(\hat{k}\times\vec{\epsilon}_{(\gamma)})]\,
Y_{MV}
\right\}
\nnb \\ &&
+ \chi_1^\dagger \sigma_2\tau_2\chi_2^{T\dagger} \,
i\vec{\epsilon}_{(d)}\cdot (\hat{k}\times \vec{\epsilon}_{(\gamma)})\,
Y_{MS}\,,\label{eq:Amplitude}
\eea
where
$\vec{\epsilon}_{(d)}$ and $\vec{\epsilon}_{(\gamma)}$ are 
spin polarization vectors for the incoming deuteron and photon,
respectively, while $\chi_1^\dagger$ and $\chi_2^\dagger$ are 
the spinors of the outgoing nucleons.
The coefficients of the terms in Eq.~(\ref{eq:Amplitude})
are given as 
\bea
X_{MV} &=& - \sqrt{\frac{\pi\gamma}{1-\gamma\rho_d}}
\frac{1}{
\frac{1}{a_0}
+ip
-\frac12r_0p^2}
\frac{1}{2m_N}
\nnb \\ && \times
\left\{
\mu_V\left[
{\rm arccos}\left(
\frac{m_N}{\sqrt{
(m_N+\frac12\omega)^2-p^2
}}
\right)
+i\ln\left(
\frac{m_N+\frac12\omega+p}{\sqrt{
(m_N+\frac12\omega)^2-p^2
}}
\right)
\right]
\right.
\nnb \\ && \left.
- \frac{\mu_V}{m_N} \left(
\frac{1}{a_0}
+ ip
-\frac12r_0 p^2
\right) F^+
+ \omega L_1
\right\}\,,
\\
X_{MS} &=& - \sqrt{\frac{\pi\gamma}{1-\gamma\rho_d}}
\frac{1}{
\gamma
%\frac{1}{a_0}
+ip
-\frac12\rho_d(\gamma^2+p^2)}
\frac{1}{2m_N}
\nnb \\ && \times
\left\{
\mu_S\left[
{\rm arccos}\left(
\frac{m_N}{\sqrt{
(m_N+\frac12\omega)^2-p^2
}}
\right)
+i\ln\left(
\frac{m_N+\frac12\omega+p}{\sqrt{
(m_N+\frac12\omega)^2-p^2
}}
\right)
\right]
\right.
\nnb \\ && \left.
- \frac{\mu_S}{m_N} \left[
\gamma %\frac{1}{a_0}
+ ip
-\frac12\rho_d(\gamma^2+p^2)
\right] F^+
+ 2 \omega L_2
\right\}\,,
\\
X_{EV} &=& \sqrt{
\frac{\pi\gamma}{1-\gamma\rho_d}
} \frac{1}{m_N^2}
\frac{p}{\omega}F^+\,,
\ \ \
Y_{ES} = \sqrt{
\frac{\pi\gamma}{1-\gamma\rho_d}
} \frac{1}{m_N^2}
\frac{p}{\omega}F^-\,,
\\
Y_{MV} &=& \sqrt{
\frac{\pi\gamma}{1-\gamma\rho_d}
} \frac{\mu_V}{2m_N^2} F^-\,,
\ \ \
Y_{MS} = \sqrt{
\frac{\pi\gamma}{1-\gamma\rho_d}
} \frac{\mu_S}{2m_N^2} F^-\,,
\eea
%where
with
\bea
2F^\pm =
\frac{1}{1 + \frac{\omega}{2m_N}
- \frac{\vec{p}\cdot\hat{k}}{m_N}}
\pm \frac{1}{1 + \frac{\omega}{2m_N}
+ \frac{\vec{p}\cdot\hat{k}}{m_N}}\,,
\eea
where $p=|\vec{p}|$, 
and $\omega$ is the incoming photon energy
in the c.m. frame. 
Low energy constant $L_1$ is fitted to reproduce the 
total cross section of a thermal neutron capture on a proton,
and $L_2$ is determined with the magnetic moment of the deuteron
\cite{ando2005}. 
The process $\gamma d\to n p$ and its inverse
process $np\to d\gamma$ share the same amplitude structure
with only changing kinematics. Keeping in mind this kinematic
difference, we can apply the transition amplitude of 
$\gamma d\to n p$ to the calculation of the amplitude of $np\to d\gamma$.
It is noteworthy that kinematics makes the $M1$ transition
as a dominant contribution to $np\to d\gamma$
while $E1$ transition to $\gamma d\to n p$.

Square of the amplitude with no polarization then reads
\bea
S^{-1}\sum_{spin} |A|^2 &=&
16\left(
|X_{MS}|^2 + |Y_{MV}|^2
\right)
+ 8\left(
|X_{MV}|^2 + |Y_{MS}|^2
\right)
\nnb \\ &&
+ 12 [1-(\hat{p}\cdot\hat{k})^2]
\left(
|X_{EV}|^2 + |Y_{ES}|^2
\right) \,,
\eea
where the symmetry factor $S$ is equal to 2 in the present case. 

\section{Results}
\subsection{Asymmetry of photon direction in 
${\vec n} p\to d \gamma$}

\begin{figure}
\begin{center}
\epsfig{file=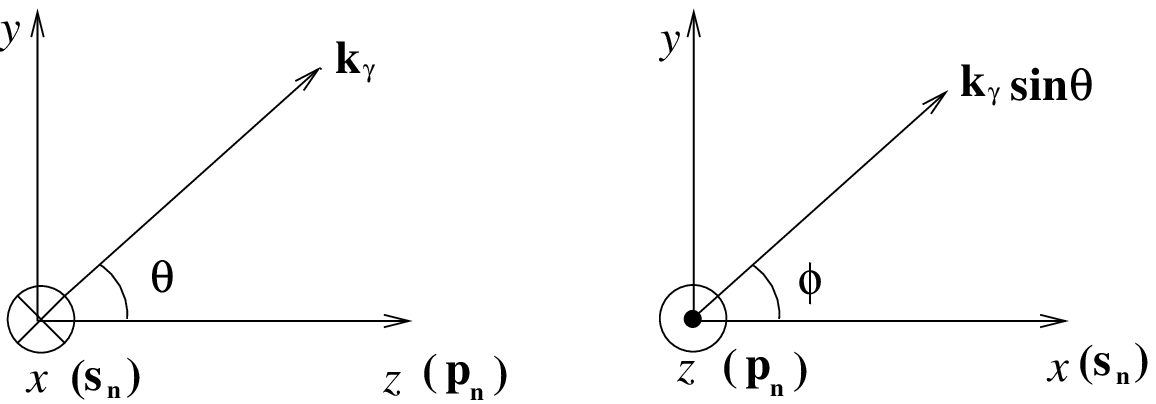,width=7.0cm}
\caption{
Coordinate system used for the calculation of $A^{LR}_\gamma$
in $\vec{n}p \to d\gamma$.
}
\label{fig:coordinate}
\end{center}
\end{figure}

Fig.~\ref{fig:coordinate} shows the coordinate system we use in
the calculation.
We are interested in the asymmetry of out-going photons 
with respect to the axix $x$,
which is proportional to $\hat{k}_\gamma \cdot (\hat{p}_n \times \hat{s}_n)
= \sin \theta \sin \phi$~\cite{gibson1997}.
Since the asymmetry is to the left and the right of the 
neutron spin, and conserves the sign under the parity conversion,
we call it parity-conserving(PC) left-right asymmetry, $A^{LR}_\gamma$.
There is another PC asymmetry, which is parallel and anti-parallel to
the neutron spin. 
Consideration of this asymmetry will be reported elsewhere \cite{liu2011}. 
Retaining the term proportional to $\sin \phi$ in the differential
cross section, we have
\bea
\frac{d\sigma}{d\Omega}=I_0(\theta)[1+P_n B(\theta)\sin\phi],
\eea
where $P_n$ is a transverse polarization of the neutron.
At low energies, $P$-wave contribution will dominate, and thus we can
approximate as $B(\theta)\simeq A^{LR}_\gamma \sin\theta$.
Then we can obtain $A^{LR}_\gamma$ with an approximate relation
\bea
\left. A^{LR}_\gamma \simeq 
\frac{\sigma_+-\sigma_-}{\sigma_++\sigma_-} \right|_{\theta=\frac{\pi}{2}},
\eea
where $\sigma_\pm$ denote the differential cross sections with
$\phi = \pm \frac{\pi}{2}$, respectively.
With low-energy neutrons, we obtain the numerical results
\bea
A^{LR}_\gamma (3\, \mbox{meV})=6.10\times 10^{-9}, \,\,
A^{LR}_\gamma (10\, \mbox{meV})=2.03\times 10^{-8}. 
\eea
Results at 3 meV obtained in SNPA are 
$A^{LR}_\gamma = 0.607 \times 10^{-8},\, 0.668 \times 10^{-8},\,
0.665 \times 10^{-8}$ with RSC, Av14 and Nijmegen93 potentials, 
respectively \cite{gibson1997}.
Our result agrees to that with RSC, but there is about 10~\% 
suppression to those with Av14 and Nijmegen93.
%
%\begin{figure}
%\begin{center}
%\epsfig{file=Agam1MevAngle.eps,width=6.0cm}
%\caption{Angle dependence of $A_\gamma^{PC}$ at 
%at $1$ MeV(solid line) and $0.5$ MeV(dashed line).
%}
%\end{center}
%\end{figure}
%
%
%\begin{figure}
%\begin{center}
%\epsfig{file=AgamEnergyDep.eps,width=6.0cm}
%\caption{Energy dependence of $A_\gamma^{PC}$
%at angle $\theta=\frac{\pi}{2}$(solid line) and  
%$\theta=\frac{\pi}{4}$(dashed line).
%}
%\end{center}
%\end{figure}
%
\subsection{Linear polarization asymmetry in ${\vec \gamma} d\to n p$}

Another polarization observable $\Sigma^l(\theta)$, the linear polarization
asymmetry in ${\vec \gamma}d\to n p$ is defined as
\bea
\frac{d\sigma}{d\Omega}&=&
\frac{d\sigma_0}{d\Omega}(1+P_l^\gamma\Sigma^l(\theta)\cos(2\phi)),\no
\Sigma^l(\theta)&=&\frac{\sigma_{||}(\theta)-\sigma_{\perp}(\theta)}{\sigma_{||}(\theta)+\sigma_{\perp}(\theta)},
\eea
where the linear polarization is parallel or perpendicular to 
the reaction plane. 
Choosing
$\hat{k}_\gamma=\hat{z}$, $\hat{p}_n=(\sin\theta,0,\cos\theta)$,
$\vec{\epsilon}_{(\gamma)}=\hat{x}$ for parallel photons, and
$\vec{\epsilon}_{(\gamma)}=\hat{y}$ for perpendicular photons,
we obtain
\bea
\Sigma^l(\theta)=\frac{3\sin^2\theta(|Y_{ES}|^2+|X_{EV}|^2)}
{4|X_{MS}|^2+2|X_{MV}|^2+4|Y_{MV}|^2+2|Y_{MS}|^2
+3\sin^2\theta(|Y_{ES}|^2+|X_{EV}|^2)}.
\eea
\begin{figure}
\begin{center}
\epsfig{file=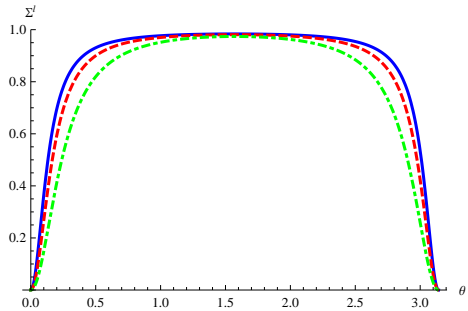,width=6.0cm}
\caption{Angle dependence of $\Sigma^l$
at $10$(solid line), $30$(dashed line), and $60$(dot-dashed line) MeV.
}
\label{fig:sigmal}
\end{center}
\end{figure}
$\Sigma^l(\theta)$'s with the incident 
photon energies 10, 30, and 60 MeV are shown in Fig.~\ref{fig:sigmal}.
Comparing the results with those reported in Ref.~\cite{evegeny2011},
the results at $\omega = 10$ MeV agree well, but there are significant
discrepancies at higher energies.

\section{Summary}

We studied spin-dependent observables for $n p\leftrightarrow d \gamma$
processes. At very low energies pionless EFT agrees with other 
theoretical calculations. Comparison with experiment
will confirm the accuracy of the theory because spin-dependent
observables are more sensitive to the amplitudes than the total cross sections.
However, our results disagree with other theoretical calculations 
at energies $E>10$ MeV. This may be explained 
by the lack of higher partial waves or higher order operators
in our calculation. Thus, more involved calculation 
is necessary.

\section*{Acknowledgments}

The work of CHH and SIA is supported by the Basic Science Research 
Program through the National Research Foundation of Korea (NRF)
funded by the Ministry of Education, Science and Technology 
(2010-0023661).
The work of YHS is supported by the US Department of Energy 
under Contract No. DE-FG02-09ER41621.
KK's work is partly supported by the US National Science Foundation
under grant number PHY-0758114.

%\begin{acknowledgements}
%\end{acknowledgements}

% BibTeX users please use
%\bibliographystyle{spbasic}


\end{document}